\documentclass[letterpaper,12pt]{article}   
\usepackage{osajnl} 
\usepackage{graphics}
\begin{document}

\title{Optical characterization of ultra-high diffraction efficiency gratings}
\author{A. Bunkowski, O. Burmeister}
\address{Max-Planck-Institut f\"ur Gravitationsphysik
 (Albert-Einstein-Institut) and\\ Universit\"at Hannover,
 Callinstr. 38, 30167 Hannover, Germany}
\email{alexander.bunkowski@aei.mpg.de}
\author{T. Clausnitzer, E.-B. Kley, and A. T\"unnermann}
\address{Institut f\"ur Angewandte Physik,
Friedrich-Schiller-Universit\"at Jena, Max-Wien-Platz 1, 07743
Jena, Germany }
\author{K. Danzmann, and R. Schnabel}
\address{Max-Planck-Institut f\"ur Gravitationsphysik
 (Albert-Einstein-Institut) and\\ Universit\"at Hannover,
 Callinstr. 38, 30167 Hannover, Germany}

\begin{abstract}We report on the optical characterization of an ultra-high
diffraction efficiency grating in 1st order Littrow configuration.
The apparatus used was an optical cavity built from the grating
under investigation and an additional high reflection mirror.
Measurement of the cavity finesse provided precise information
about the grating's diffraction efficiency and its optical loss.
We measured a finesse of 1580 from which we deduced a diffraction
efficiency of (99.635$\pm$0.016)\,\% and an overall optical loss
due to scattering and absorption of just 0.185\,\%.
Such high quality gratings, including the tool used for their
characterization, might apply for future gravitational wave
detectors.
For example the demonstrated cavity itself presents an
all-reflective, low-loss Fabry-Perot resonator that might replace
conventional arm cavities in advanced high power Michelson
interferometers.
\end{abstract}

\maketitle 
High quality optics are key devices in laser interferometric
precision measurements.
Especially for high power laser applications with nested cavities,
such as in gravitational wave detectors\cite{GWinterferometer04},
mirrors with high reflectivity and low overall optical loss are
essential.
Mirrors with a power reflectance greater than 99.9998\,\% for a
given laser wavelength have been reported \cite{Rempe}.
The overall optical loss consisting of stray light from the
surface, transmission, and absorption in the coating was as low as
1.6\,~ppm\cite{Rempe}.

Gratings are traditionally used in applications where one wants to
spatially resolve different optical wavelengths, e.g. in
spectrographs or pulse compressors/stretchers for short pulse
laser systems.
In these applications high diffraction efficiency over a range of
optical wavelengths is desired.
Dielectric reflection gratings having diffraction efficiencies of
96\,\%, 97\,\%, and 99\,\% have been
reported\cite{Perry,Hehl,Destouches}.
However the measurement techniques used there only allowed for a
rough estimation of the diffraction efficiency and no error bars
for the values were given.
Diffraction gratings may also be used in advanced high power laser
interferometers\cite{Drever96,Sun97}, where they allow for the
all-reflective realization of beam splitters and cavity couplers
and therefore may help to reduce thermal effects in the substrate
like thermal lensing\cite{Winkler} and thermo-refractive
noise\cite{Braginsky}.
In interferometric applications monochromatic laser light is used
and the wavelength dispersive property of the gratings is not
essential.
The point of interest lies in the number and the properties of the
reflective diffraction ports  and their couplings that determine
the interference between input beams.
Two different all-reflective resonator concepts have been
demonstrated to date.
High efficiency gratings in first-order Littrow configuration form
cavity couplers with two ports analogous to partially transmitting
mirrors\cite{Sun97}.
Low efficiency gratings in second-order Littrow configuration can
be used as low-loss couplers with three ports\cite{Bunkowski}.
Analogous to conventional mirrors however, optical loss in terms
of scattering or absorption has to be minimized in order to gain
maximum laser power build-up and measurement sensitivity.
The question therefore arises if high-efficiency gratings with
highly corrugated surfaces will ever be able to fulfill the strict
low scattering loss requirements.

In this article we report on the optical characterization of a
high efficiency grating in view of applications in interferometry.
The grating was used in first-order Littrow configuration to
couple laser light into a Fabry-Perot cavity with a measured
finesse of $1580\pm60$.
This experiment allowed for the accurate measurement of both the
grating's loss and the diffraction efficiency.
The latter one was determined to be (99.635$\pm$0.016)\,\%.
To our knowledge this is the highest and most accurately
determined value reported in the literature.

\begin{figure}[htb]
   \centerline{\includegraphics[width=8cm]{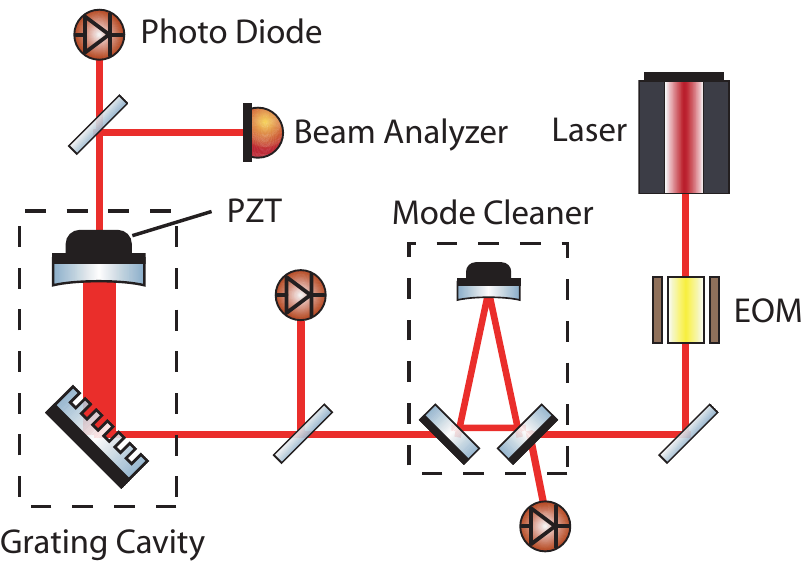}}
    \caption{(Color online) Experimental setup; EOM: electro optical modulator;
    PZT: cavity mirror with piezoelectric transducer for length control.
    \label{setup4}}
\end{figure}

The grating device was designed for a laser wavelength  of
1064\,nm and a Littrow angle of approximately $30^{\circ}$.
The grating structure had rectangular grooves with a period of
1060\,nm.
For fabrication we used electron beam direct writing (electron
beam writer LION LV1 from Leica Microsystems Jena GmbH) and
reactive ion beam etching into the top layer of a highly
reflective dielectric multilayer stack.
The stack consisted of 36 alternating layers of 195\,nm SiO$_2$
and 136\,nm Ta$_2$O$_5$ placed on a fused silica substrate with a
surface flatness of $\lambda/10$.
For the theoretical optimization of the grating we used the
rigorous coupled wave analysis\cite{Moharam}.
In order to ensure a good reproducibility and homogeneity over the
whole grating area an important concern of the design was a large
groove parameter tolerance of the diffraction efficiency.
By using SiO$_2$ with a thickness of $1.12\,\mu$m as the top layer
of the dielectric stack the theoretical design exhibited a
diffraction efficiency of more than 99\,\% for  groove depths
between 700\,nm and 850\,nm and groove widths between 530\,nm and
760\,nm.
A schematic of our experiment is seen in Fig.~\ref{setup4}. The
light source used was a 1.2 W diode pumped Nd:YAG laser at
1064\,nm (Model Mephisto from Innolight GmbH).
Before the s-polarized laser beam was sent into the grating cavity
it was spatially filtered with a triangular ring cavity (mode
cleaner)\cite{Willke}.
The highly reflective end mirror of the grating cavity was mounted
on a piezoelectric transducer (PZT) which was used to either scan
or to actively control the cavity length.
The error signal for the electronic servo loop was obtained from
the cavity transmission demodulated at the phase modulation
frequency introduced by the EOM.

In first order Littrow configuration only two diffraction orders
exist and the grating (subscript 1) is characterized by the zeroth
and first order amplitude diffraction efficiencies $r_1$ and
$\eta_1$, respectively, as well as the loss amplitude $l_1$.
Similarly the cavity end mirror (subscript 2) is described by
$r_2$, $t_2$ and $l_2$.
Energy conservation implies
\begin{eqnarray}\label{eq:g1}
    r_1^2 + \eta_1^2 + l_1^2 &=&1,\\
    r_2^2 + t_2^2    + l_2^2 &=&1\, .
    \label{eq:m2}
\end{eqnarray}

\begin{figure}[htb]
  \centerline{\includegraphics[width=8cm]{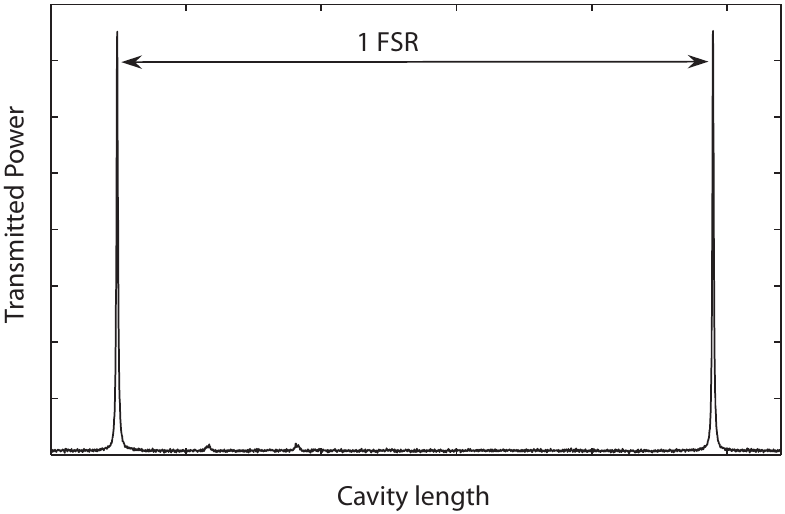}} 
    \caption{Transmitted power of the cavity: Photo diode signal
    behind the grating cavity as the cavity is linearly scanned
    over one free spectral range (FSR).
    \label{fsr}}
\end{figure}

Figure~\ref{fsr} shows the transmission spectrum of the cavity as
the PZT is linearly scanned over one free spectral range of the
cavity.
In addition to the peaks of the fundamental mode of the cavity
there are only two samller peaks from higher order modes visible
indicating a good matching of laser beam and cavity mode.

\begin{figure}[htb]
    \centerline{\includegraphics[width=8cm]{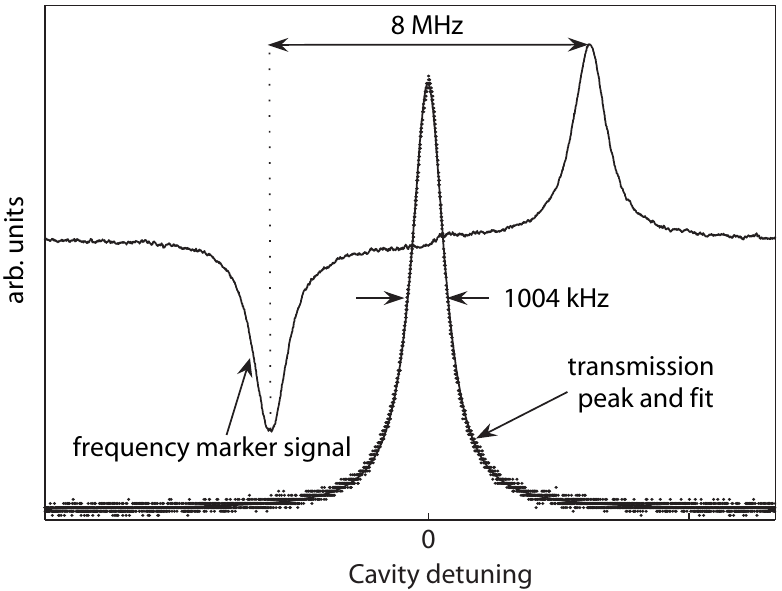}}
    \caption{Scan over one cavity transmission peak.
    The x-axis was calibrated with
    $\pm 4$MHz marker signals.
    \label{Fig3fwhm}}
\end{figure}

A method to obtain a precise value for a mirror reflectance close
to unity is a measurement of the finesse $F$ of a cavity
consisting of a mirror with known reflectance and the one in
question.
For the first time we applied this method to characterize a  high
efficiency grating.
If losses due to absorption in the space between the mirrors
(which would appear additionally to $l_1$ and $l_2$) are neglected
the finesse $F$ of a two mirror Fabry-Perot resonator depends on
the reflectance of the two end mirrors only.
In our case one of the end mirrors is a grating and the finesse
can be approximated by
\begin{equation} \label{eq:finesse}
    F=\pi(\eta_1 r_2)^{1/2}/(1-\eta_1 r_2).
\end{equation}
For a cavity of length L its free spectral range is given by
$f_{\mathrm{FSR}}=c/2L$ where $c$ is the speed of light.
The ratio of $f_{\mathrm{FSR}}$ to the full width at half maximum
$f_{\mathrm{FWHM}}$ of the Airy transmission spectrum peaks
determines the finesse
\begin{equation} \label{eq:finesse2}
    F=f_{\mathrm{FSR}}/f_{\mathrm{FWHM}}.
\end{equation}
The length of the cavity was measured to $L=(94 \pm 1)\,$mm,
resulting in $f_{\mathrm{FSR}}\approx 1.6\,$GHz.
The cavity linewidth was measured utilizing frequency marker
signals.
The laser light was phase modulated at $f_{\mathrm{mod}}=4\,$MHz
using an electro optical modulator (EOM).
The AC output of the photodiode in front of the grating cavity was
then demodulated at $f_{\mathrm{mod}}$.
For the correct demodulation phase this signal shows a minimum and
a maximum at exactly $\pm f_{\mathrm{mod}}$ and can be used to
calibrate the x-axis in Fig.~\ref{Fig3fwhm}.
The figure shows a typical measured DC signal of the photo diode
behind the cavity as well as the marker signals at
$f_{\mathrm{marker}}=\pm (4 \pm 0.04)$MHz while the cavity was
linearly scanned with 1\,kHz repetition rate.
The uncertainty in the position of the marker signal is due to an
error in the demodulation phase.
A fit of the transmission signal to the well known Airy function
of cavities permitted the calculation of the width of the
transmission peak.
Due to nonlinearities in the piezoelectric transducer and acoustic
vibrations there is a statistical variation of the linewidth of
the peak.
We averaged over 75 measurement using different operating points
of the piezoelectric transducer and could reduce the statistical
error in the peak width to $\pm3.5\%$.

With Eq.~(\ref{eq:finesse2}) we could calculate the finesse of the
cavity  $F=1580 \pm 60$.
The cavity end mirror was super-polished and coated by REO
(Research Electro-Optics, Inc) and specified to have values of
$t^2_2=300\pm30$ ppm and $l^2_2< 30$\,ppm.
From these specifications we estimate the mirror's reflectivity to
$r_2^2=(99.9685 \pm 0.0034)\%$.
With Eqs.~(\ref{eq:m2}) and (\ref{eq:finesse}) we obtained
$\eta_1^2=(99.635\,\pm0.016)\,\%$ for the grating's 1st order
diffraction efficiency.
The error in $\eta_1^2$ results from an error propagation of each
known uncertainty of the quantities $L, f_{\mathrm{marker}},$
fitted peak width and $r_2^2$ as shown in table~1.
\begin{table}[h]
{\bf \caption{Error propagation}}\begin{center}
\begin{tabular}{c c c}
\hline
Quantity & error & proj. error for $\eta_1^2$ [ppm] \\
\hline
L& $\pm1$\,mm & $\pm 48$\\
$f_{\mathrm{marker}}$ & $\pm 40$\,kHz & $\pm 43$\\
peak width& $\pm 3.5\,\%$ & $\pm 143$ \\
$r_2^2$ & $\pm 34$\,ppm & $\pm 34$ \\ \hline
\multicolumn{2}{c}{Total RMS error expected} & $\pm 160$
\\ \hline
\end{tabular}
\end{center}
\end{table}

The specular reflection of the grating was measured independently
with a calibrated power meter to be $r^2_1=(0.18\,\pm0.009)\,\%$.
Hence we calculated the overall loss of the grating according to
Eq.~(\ref{eq:g1}) to be $l^2_1=(0.185\pm 0.025)\,\%$.
We emphasize that this loss contained all contributions from
scattering, absorption, transmission, and higher diffraction
orders.
To our best knowledge this result presents the lowest and most
accurately determined grating loss reported in the literature.
Previous results were those by Perry \emph{et al.}\cite{Perry} and
Hehl \emph{et al.}\cite{Hehl} who reported 1.5\,\% and 1 - 2\,\%
loss, respectively.
Destouches \emph{et. al.}\cite{Destouches} have not commented on
the loss.

\begin{figure}[htb]
    \centerline{\includegraphics[width=8.5cm]{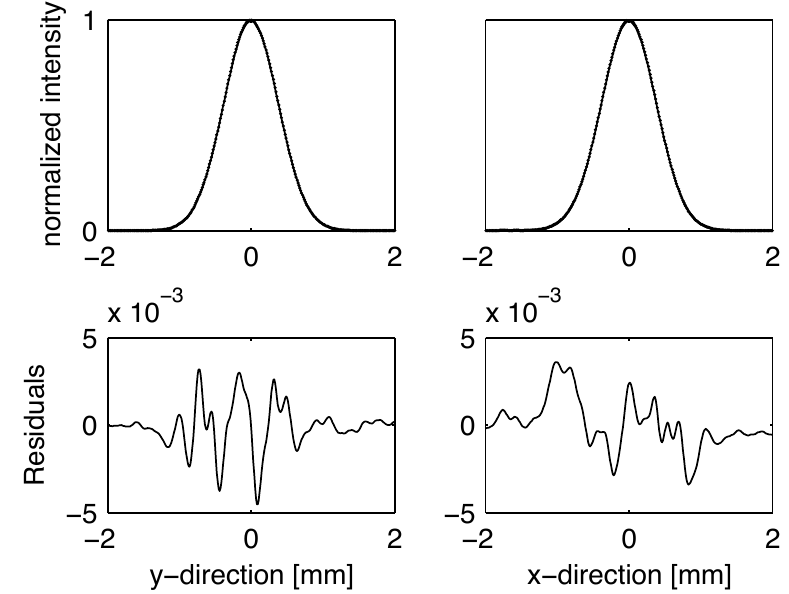}}
    \caption{Spatial beam profile of the laser beam after the cavity
    for horizontal (perpendicular to the grating lines) and vertical (parallel to the grating lines) direction. Top:
    measured points (dots) and best gaussian fit (solid line);
    Bottom: Residuals between measurement and fit.}
    \label{Fig4}
\end{figure}

In addition to the grating's loss we also investigated its
influence on the laser beam's spatial profile.
Again a cavity in first order Littrow configuration was set up
with cavity mode waist on the gratings surface now using an end
mirror with power reflectivity $r^2_2=99\,\%$ to reduce the
finesse value and to increase transmission.
The cavity length was locked to one of the transmission maxima
using the radiofrequency phase-modulation technique described
above.
The beam profile for the horizontal and vertical directions were
measured after the cavity using a seven-blade tomographic profiler
(SuperBeamAlyzer from Melles Griot) and fitted with a Gaussian
model, as shown in Fig.~\ref{Fig4}. The sum of the absolute
differences between the value of every measured point and the
fitted function divided by the sum of the values of all fitted
points is a measure of how much beam power can be represented by a
gaussian function.
For both directions we obtained values of greater than 99\%.
For this experiment the modecleaner had been taken out which
allowed us to observe a modecleaning effect from the grating
cavity.
We characterized the laser beam behind the EOM using the same
apparatus and got spatial profiles that were described by a
gaussian function by only 98\,\%.

In summary we presented a detailed characterization of diffraction
efficiency and over-all loss of a grating in 1st order Littrow
mount.
The grating's diffraction efficiency showed an outstanding high
value which enabled the construction of a high finesse cavity as a
characterizing tool.
The value of the finesse was limited by the first order
diffraction efficiency. This is in contrast to Ref.
\cite{Bunkowski} where a low diffraction efficiency grating was
characterized with a high finesse cavity and where the limit for
the finesse was given by the specular reflectivity of the grating.
Our approach is a valuable diagnostic tool to improve future
techniques of grating fabrication, since all types of loss are
simultaneously detected.
We expect that with improved technology high grating efficiencies
with simultaneously low loss are possible that even fulfill the
strict requirements of future interferometers such those for
gravitational wave detection.

This work was supported by the Deutsche Forschungsgemeinschaft
within the Sonderforschungsbereich TR7.


\end{document}